\documentclass[aps,prb,twocolumn,showpacs,superscriptaddress]{revtex4}

\usepackage{graphicx}
\usepackage{amsmath}
\usepackage{amsfonts}
\usepackage{amssymb}
\usepackage{latexsym}
\usepackage{dcolumn}

\newcommand{\mg}{\textrm{M\lowercase{g}}}
\newcommand{\tm}{\textrm{TM}}

\newcommand{\hyd}{\textrm{H}_{2}}
\newcommand{\mgh}{\textrm{M\lowercase{g}}\textrm{H}_{2}}
\newcommand{\sch}{\textrm{S\lowercase{c}}\textrm{H}_{2}}
\newcommand{\tih}{\textrm{T\lowercase{i}}\textrm{H}_{2}}
\newcommand{\vh} {\textrm{V}\textrm{H}_{2}}
\newcommand{\crh}{\textrm{C\lowercase{r}}\textrm{H}_{2}}

\newcommand{\tmh}{\textrm{TM}\textrm{H}_{2}}

\newcommand{\mgtmha}{\textrm{M\lowercase{g}}_{x}\textrm{TM}_{(1-x)}\textrm{H}_{2}}
\newcommand{\mgtma}{\textrm{M\lowercase{g}}_{x}\textrm{TM}_{(1-x)}}
\newcommand{\mgtmhb}{\textrm{M\lowercase{g}}_{0.75}\textrm{TM}_{0.25}\textrm{H}_{2}}
\newcommand{\mgtmb}{\textrm{M\lowercase{g}}_{0.75}\textrm{TM}_{0.25}}

\newcommand{\mgscha}{\textrm{M\lowercase{g}}_{x}\textrm{S\lowercase{c}}_{(1-x)}\textrm{H}_{2}}
\newcommand{\mgsca}{\textrm{M\lowercase{g}}_{x}\textrm{S\lowercase{c}}_{(1-x)}}

\newcommand{\mgscb}{\textrm{M\lowercase{g}}_{0.75}\textrm{S\lowercase{c}}_{0.25}}
\newcommand{\mgschy}{\textrm{M\lowercase{g}}_{x}\textrm{S\lowercase{c}}_{(1-x)}\textrm{H}_{y}}

\newcommand{\mgtiha}{\textrm{M\lowercase{g}}_{x}\textrm{T\lowercase{i}}_{(1-x)}\textrm{H}_{2}}
\newcommand{\mgtia}{\textrm{M\lowercase{g}}_{x}\textrm{T\lowercase{i}}_{(1-x)}}

\newcommand{\mgtib}{\textrm{M\lowercase{g}}_{0.75}\textrm{T\lowercase{i}}_{0.25}}
\newcommand{\mgtihy}{\textrm{M\lowercase{g}}_{x}\textrm{T\lowercase{i}}_{(1-x)}\textrm{H}_{y}}

\newcommand{\mgvha} {\textrm{M\lowercase{g}}_{x}\textrm{V}_{(1-x)}\textrm{H}_{2}}
\newcommand{\mgva} {\textrm{M\lowercase{g}}_{x}\textrm{V}_{(1-x)}}
\newcommand{\mgvhb} {\textrm{M\lowercase{g}}_{0.75}\textrm{V}_{0.25}\textrm{H}_{2}}
\newcommand{\mgvb} {\textrm{M\lowercase{g}}_{0.75}\textrm{V}_{0.25}}

\newcommand{\mgcrha}{\textrm{M\lowercase{g}}_{x}\textrm{C\lowercase{r}}_{(1-x)}\textrm{H}_{2}}
\newcommand{\mgcra}{\textrm{M\lowercase{g}}_{x}\textrm{C\lowercase{r}}_{(1-x)}}
\newcommand{\mgcrhb}{\textrm{M\lowercase{g}}_{0.75}\textrm{C\lowercase{r}}_{0.25}\textrm{H}_{2}}
\newcommand{\mgcrb}{\textrm{M\lowercase{g}}_{0.75}\textrm{C\lowercase{r}}_{0.25}}

\begin{document}

\title{Tunable Hydrogen Storage in Magnesium - Transition Metal Compounds}
\author{S\"{u}leyman Er}
\affiliation{Computational Materials Science, Faculty of Science and Technology and MESA+
Research Institute, University of Twente, P.O. Box 217, 7500 AE Enschede, The Netherlands}
\author{Dhirendra Tiwari}
\affiliation{Computational Materials Science, Faculty of Science and Technology and MESA+
Research Institute, University of Twente, P.O. Box 217, 7500 AE Enschede, The Netherlands}
\author{Gilles A. de Wijs}
\affiliation{Electronic Structure of Materials, Institute for Molecules and Materials, Faculty
of Science, Radboud University Nijmegen, Heyendaalseweg 135, 6525 AJ Nijmegen, The Netherlands}
\author{Geert Brocks}
\affiliation{Computational Materials Science, Faculty of Science and Technology and MESA+
Research Institute, University of Twente, P.O. Box 217, 7500 AE Enschede, The Netherlands}

\date{\today}

\begin{abstract}
Magnesium dihydride ($\mgh$) stores $7.7$ weight $\%$ hydrogen, but it suffers from a high
thermodynamic stability and slow (de)hydrogenation kinetics. Alloying Mg with lightweight transition metals (TM = Sc, Ti, V, Cr) aims at improving the thermodynamic and kinetic properties. We study the structure and stability of Mg$_x$TM$_{1-x}$H$_2$ compounds, $x=[0$-$1$], by first-principles calculations at the level of density functional theory. We find that the experimentally observed sharp decrease in hydrogenation rates for $x\gtrsim0.8$ correlates with a phase transition of Mg$_x$TM$_{1-x}$H$_2$ from a fluorite to a rutile phase. The stability of these compounds decreases along the series Sc, Ti, V, Cr.  Varying the transition metal (TM) and the composition $x$, the formation enthalpy of Mg$_x$TM$_{1-x}$H$_2$ can be tuned over the substantial range $0-2$ eV/f.u. Assuming however that the alloy Mg$_x$TM$_{1-x}$ does not decompose upon dehydrogenation, the enthalpy associated with reversible hydrogenation of compounds with a high magnesium content ($x=0.75$) is close to that of pure Mg.
\end{abstract}
\pacs{71.20.Be, 71.15.Nc, 61.66.Dk, 61.50.Lt}
\maketitle

\section{Introduction}
Hydrogen is a clean energy carrier and an alternative to carbon based fuels in the long run. \cite{coontz2004nss} Mobile applications require a compact, dense and safe storage of hydrogen
with a high-rate loading and unloading capability. \cite{zuttel2003mhs,zuttel2004hsm} Lightweight
metal hydrides could satisfy these requirements. \cite{schlapbach2001hsm,bogdanovic2003ihs} Metal
hydrides are formed by binding hydrogen atoms in the crystal lattice, resulting in very high
volumetric densities. Reasonable hydrogen gravimetric densities in metal hydrides can be achieved
if lightweight metals are used.

$\mgh$ has been studied intensively since it has a relatively high hydrogen gravimetric density of
$7.7$ wt. $\%$. Bottlenecks in the application of $\mgh$ are its thermodynamic stability and slow (de)hydrogenation kinetics. These lead to excessively high operating
temperatures ($573-673$ K) for hydrogen release. \cite{stampferjr1960mhs,huot2001mam,grochala2004tdn}                 The hydrogen (de)sorption rates can be improved by decreasing the particle size down to nanoscales. \cite{zaluska2001sca,dornheim2006ths,li2007mne} It is predicted that particles smaller than $1$ nm have a markedly decreased hydrogen desorption enthalpy, which would lower the operating temperature.\cite{wagemans2005hsm} The production of such small particles is nontrivial, however, and the hydrogen (de)sorption rates of larger nanoparticles are still too low.

An additional way of improving the (de)hydrogenation kinetics of $\mgh$ is to add transition
metals (TMs).\cite{zaluska2001sca,pelletier2001hdm,vonzeppelin2002hdk,yao2006:jpcb} Usually only a few wt. \% is added, since TMs are thought to act as catalysts for the dissociation of hydrogen
molecules. Recently however, Notten and co-workers have shown that the (de)hydrogenation kinetics
is markedly improved by adding more TM and making alloys $\mgtma$, TM$=$Sc, Ti, $x\lesssim 0.8$.
\cite{notten2004hed,niessen2005ehs,niessen2005hst, kalisvaart2006ehs,
niessen2006epc,vermeulen2006hsm,
borsa2006mth,borsa2007soa,vermeulen2007ter,kalisvaart2007mtb,gremaud2007hoc} The basic ansatz is
that the rutile crystal structure of $\mgh$ enforces an unfavorably slow diffusion of
hydrogen atoms. \cite{buschow1982hfi} $\sch$ and $\tih$ have a fluorite structure, which would be more favorable for fast hydrogen kinetics. By adding a sufficiently large fraction of these TMs one could force the $\mgtma$ compound to adopt the fluorite structure.

In this paper we examine the structure and stability of $\mgtmha$, TM = Sc, Ti, V, Cr, compounds by first-principles calculations. In particular, we study the relative stability of the rutile versus the fluorite structures. This paper is organized as follows. In Sec.~\ref{sec:computational} we discuss the computational details. The calculations are benchmarked on the $\tmh$ simple hydrides. The structure and formation enthalpies of the compounds $\mgtmha$ are studied in Sec.~\ref{sec:results} and an analysis of the electronic structure is given. We discuss the hydrogenation enthalpy of the compounds in Sec.~\ref{sec:discussion} and summarize our main results in Sec.~\ref{sec:summary}.

\section{Computational Methods and Test Calculations}\label{sec:computational}
We perform first-principles calculations at the level of density functional theory (DFT) with the
PW91 functional as the generalized gradient approximation (GGA) to exchange and correlation. \cite{perdew1996gga} As transition metals have partially filled 3$d$ shells we include spin polarization and study ferromagnetic and simple antiferromagnetic orderings where appropriate.
A plane wave basis set and the projector augmented wave (PAW) formalism are used, \cite{blochl1994paw,kresse1999upp} as implemented in the VASP code. \cite{kresse1993aim,kresse1996eis} The cutoff kinetic energy for the plane waves is set
at $650$ eV. Standard frozen core potentials are applied for all the elements, except for Sc,
where we include 3\textit{s} and 3\textit{p} as valence shells, in addition to the usual 4\textit{s} and 3\textit{d} shells. The Brillouin zone (BZ) is integrated using a regular
$\textbf{k}$-point mesh with a spacing $\sim 0.02$ \AA$^{-1}$ and the Methfessel-Paxton scheme
with a smearing parameter of $0.1$ eV. \cite{methfessel1989hps}
The self-consistency convergence criterion for the energy difference between two consecutive
electronic steps is set to 10$^{-5}$ eV. Structural optimization is assumed to be complete when
the total force acting on each atom is smaller than $0.01$ eV/\upshape{\AA}. The volumes of the unit cells are relaxed, and, where appropriate, also their shapes. Finally, we calculate
accurate total energies for the optimized geometries using the linear tetrahedron method.
\cite{blochl1994itm}

To calculate the formation enthalpies of the metal hydrides we consider the following reaction.
\begin{equation}
\label{MHreaction}
     x\mg  +  (1-x)\tm  +  \hyd(g) \longrightarrow  \mgtmha.
\end{equation}
The formation enthalpies (at $T=0$) are then obtained by subtracting the total energies of the reactants from that of the product. A cubic box of size 10 \AA\ is applied for the $\hyd$ molecule. The calculated H$-$H bond length, binding energy, and vibrational frequency are $0.748$ \upshape{\AA}, $-4.56$ eV and $4351$ cm$^{-1}$, respectively, in good agreement with the experimental values of $0.741$ \upshape{\AA}, $-4.48$ eV and $4401$ cm$^{-1}$. \cite{hubher1979cdm,codata1989kvt}

Since hydrogen is a light element, the zero point energy (ZPE) due to its quantum motion, is not negligible. We find that the correction to the reaction enthalpies of Eq.~(\ref{MHreaction}) resulting from the ZPEs, is  $0.15 \pm 0.05$ eV/H$_2$, as function of the composition $x$ and the transition metal TM. Since we are mainly interested in
relative formation enthalpies, we omit the ZPE energy correction in the following.

\begin{figure}[tb]
    \centering
        \includegraphics[width=4cm]{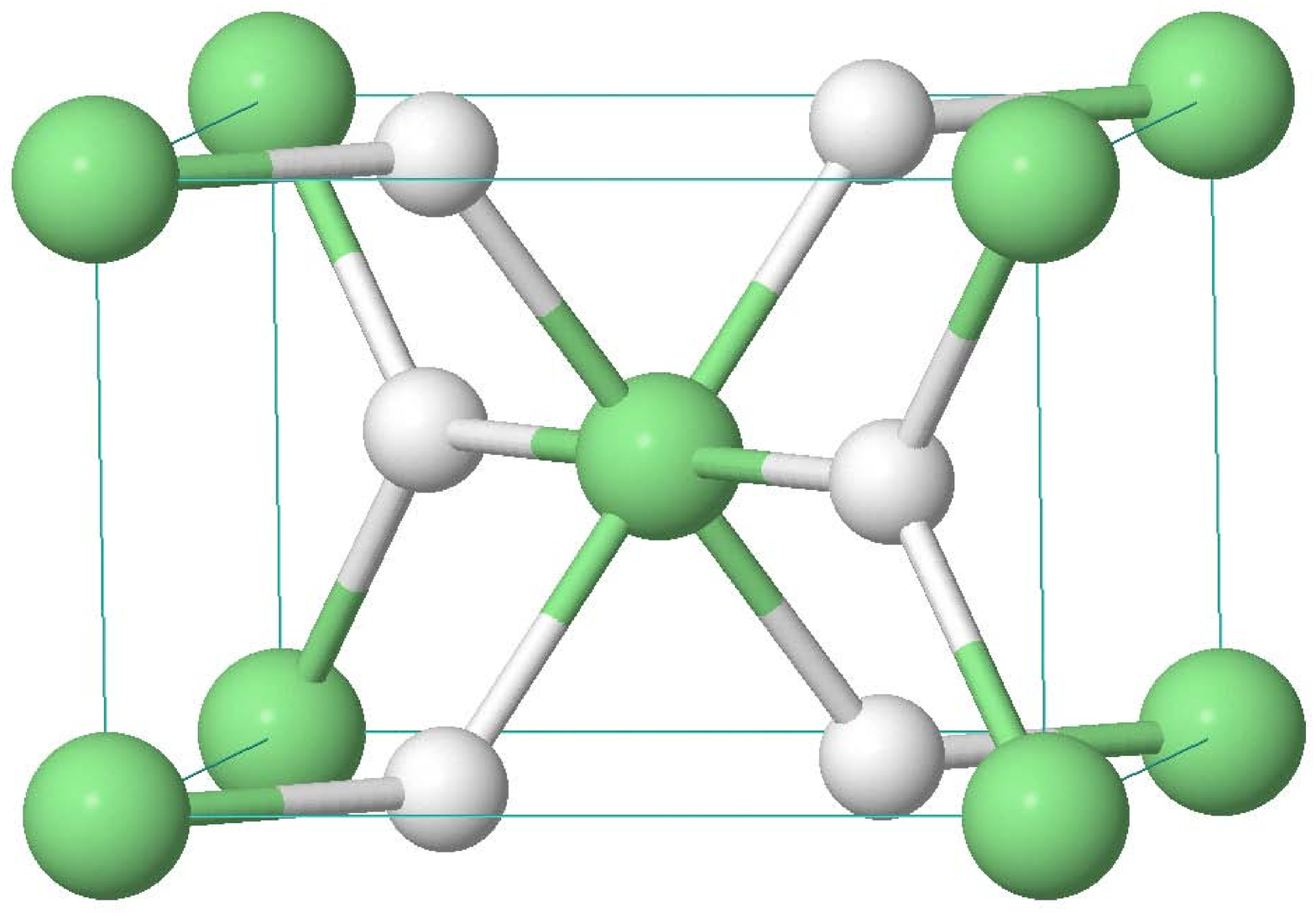}
        \includegraphics[width=4cm]{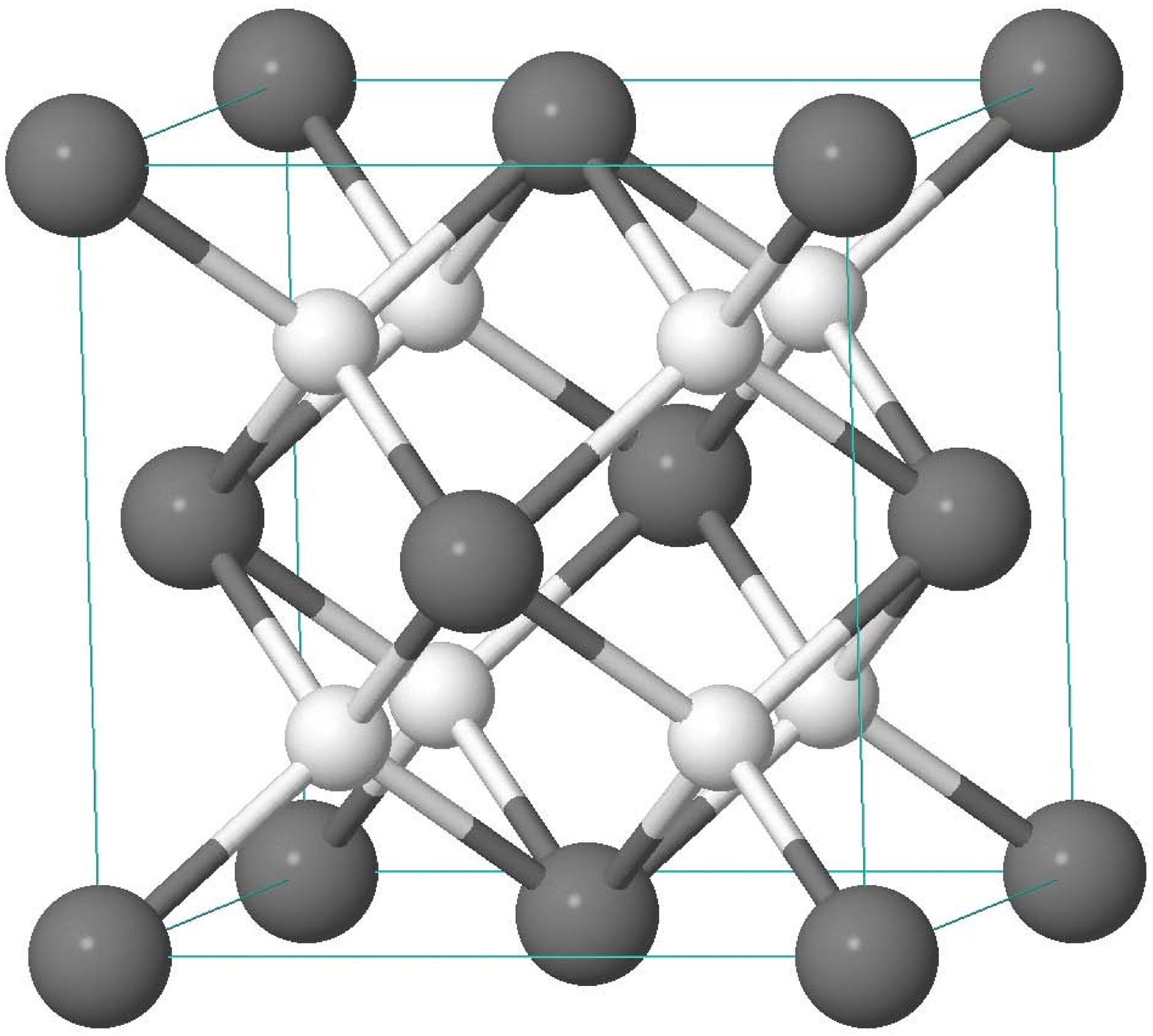}
        \caption{(Color online) (Left) rutile crystal structure of $\alpha$-$\mgh$, and (right) fluorite crystal structure of $\alpha$-$\tmh$. The white spheres represent the hydrogen atoms.}
    \label{fig:binarydihydrides}
\end{figure}

Before discussing the $\mgtmha$ compounds we benchmark our calculations on the simple compounds
$\mgh$ and $\tmh$. Under standard conditions magnesium-hydride has the rutile
structure, $\alpha$-$\mgh$, see Fig. \ref{fig:binarydihydrides}, with space group P4$_{2}$/mnm (136) and Mg and H
atoms in the 2$\textit{a}$ and 4$\textit{f}$ ($x=0.304$) Wyckoff positions, respectively. Each Mg atom is
coordinated octahedrally by H atoms, with two Mg-H distances of $1.94$ \upshape{\AA} and four
distances of $1.95$ \upshape{\AA}. First row early transition metal hydrides
crystallize in the fluorite structure, $\alpha$-$\tmh$, see Fig. \ref{fig:binarydihydrides}, with space group
Fm$\overline{3}$m (225) and TM atoms in $\textit{a}$ and H atoms in 8$\textit{c}$ Wyckoff positions.
Each TM has a cubic surrounding of H atoms with calculated TM-H bond lengths of $2.07, 1.92, 1.82$
and $1.79$ \upshape{\AA} for Sc, Ti, V, and Cr, respectively. By breaking the cubic symmetry by
hand and reoptimizing the geometry we have confirmed that the fluorite structure indeed represents a stable minimum.

The optimized cell parameters and the calculated formation enthalpies of the simple hydrides are
given in Table \ref{BHstr}. The structural parameters are in good agreement both with available
experimental data and with previous DFT calculations. The formation enthalpies of $\mgh$ and $\vh$
are somewhat underestimated by the calculations, whereas those of $\sch$ and $\tih$ are in
excellent agreement with experiment. $\crh$ is predicted to be unstable with respect to
decomposition.

\begin{table}[tb]
\centering
\caption{Optimized cell parameters $a$ $(c)$, and calculated formation enthalpies $E_f$, of elemental dihydrides in their most stable ($\alpha$) forms. All $\tmh$ have a fluorite structure, space group Fm$\overline{3}$m (225), whereas $\mgh$ has a rutile structure, space group P4$_{2}$/mnm (136). }
\label{BHstr}
\begin{ruledtabular}
\begin{tabular}{ccccc}
Compound        & \multicolumn{2}{c}{$a$ $(c)$ \AA} & \multicolumn{2}{c}{$E_f$ (eV/f.u.)} \\
                &    Calc          & Exp                            &   Calc      &  Exp\\
\hline
$\mgh$          &   4.494 (3.005)  &  4.501 (3.010)\footnotemark[1] &   $-0.66$   & $-0.76$ \\
$\sch$          &   4.775          &  4.78\footnotemark[2]          &   $-2.09$   & $-2.08$ \\
$\tih$          &   4.424          &  4.454\footnotemark[3]         &   $-1.47$   & $-1.45$ \\
$\vh$           &   4.210          &  4.27\footnotemark[4]          &   $-0.65$   & $-0.79$ \\
$\crh$          &   4.140          &  3.861\footnotemark[4]         &   $+0.13$\footnotemark[5]   &   -     \\
\end{tabular}
\end{ruledtabular}
\footnotetext[1]{Ref. \onlinecite{bortz1999shp}}
\footnotetext[2]{Ref. \onlinecite{mueller1968mh}}
\footnotetext[3]{Ref. \onlinecite{villars1991psh}}
\footnotetext[4]{Ref. \onlinecite{snavely1949ucd}}
\footnotetext[5]{Antiferromagneticly ordered.}
\end{table}

\section{Results $\mgtmha$}\label{sec:results}
\subsection{Structures and formation enthalpies}\label{sec:forment}
$\mgtmha$ has the fluorite structure for $x=0$, and the rutile structure for $x=1$. We want to
establish which of the two structures is most stable at intermediate compositions $x$. First we
summarize the current status of the experimental work on $\mgtma$ alloys.

Experimentally it has been demonstrated that $\mgsca$ alloys can be reversibly hydrogenated, both
in thin films, as well as in bulk form. \cite{notten2004hed,niessen2005ehs,niessen2006epc,kalisvaart2006ehs,latroche2006csm,magusin2008hsa} Mg and Ti do not form a stable bulk alloy, but thin films of $\mgtia$ have been made, which are
readily and reversibly hydrogenated. \cite{vermeulen2006hsm,vermeulen2006edt,borsa2006mth,borsa2007soa,vermeulen2007ter,kalisvaart2007mtb,gremaud2007hoc} Thin films of $\mgva$ and $\mgcra$ can also be easily hydrogenated. \cite{niessen2005ehs} Attempts
to produce non-equilibrium bulk $\mgtia$ alloys by ball milling of Mg and Ti or their hydrides
have had a limited success so far. \cite{liang1999cet,bobet2000sma,liang2003smt,choi2008hsp}
However, Mg$_7$TiH$_{y}$ crystals have been made using a high pressure anvil technique. \cite{kyoi2004ntm}            The same technique has been applied to produce the hydrides Mg$_6$VH$_{y}$
and Mg$_3$CrH$_{y}$. \cite{kyoi2003fmc,kyoi2004nmv,ronnebro2004scm}

The crystal structure of $\mgschy$ and $\mgtihy$ in thin films, $x\lesssim 0.8$, $y\approx 1$-2, is
cubic, with the Mg and TM atoms at fcc positions. No detectable regular ordering of Mg and TM
atoms at these positions has been found. \cite{latroche2006csm,borsa2007soa,magusin2008hsa}
In contrast, the Mg and TM atoms form simple ordered structures in the high pressure phases. \cite{kyoi2003fmc,kyoi2004ntm,kyoi2004nmv,ronnebro2004scm,ronnebro2005hsa} The hydrogen atoms in
Mg$_{0.65}$Sc$_{0.35}$H$_y$, $y\approx 1$-2, assume tetrahedral interstitial positions, as is
expected for the fluorite structure. \cite{latroche2006csm,magusin2008hsa} In the Mg$_7$TiH$_{16}$ high
pressure phase the metal atoms are in fcc positions and are ordered as in the
Ca$_7$Ge structure. \cite{kyoi2004ntm} The H atoms are in interstitial sites, but displaced from
their ideal tetrahedral positions. \cite{ronnebro2005hsa}

The latter structure can be used as a starting point to construct simple, fluorite-type structures
for $\mgtmha$, $0<x<1$. For $x=0.125, 0.875$ we use the Ca$_7$Ge structure to order the metal
atoms, for $x=0.25, 0.75$ the Cu$_3$Au ($L1_2$) structure, and for $x=0.5$ the CuAu ($L1_0$)
structure. The H atoms are placed at or close to tetrahedral interstitial positions. To model $\mgtmha$ in
rutile-type structures we use the $\alpha$-MgH$_2$ structure as a starting point. For each
composition $x$ we choose the smallest supercell of that structure corresponding to that
composition. As the atomic volumes of the various TM atoms differ, in each of the structures and
compositions the cell parameters are optimized, as well as the positions of all atoms within the
cell. Care is taken to allow for breaking the symmetry in the atomic positions. In particular the
hydrogen atoms are often displaced from their ideal tetrahedral positions. Although these structures are then no longer ideal fluorite structures anymore, we still use the term fluorite in the following. Test calculations show that the total energies of the hydrides of (quasi-)random alloys are similar to those of the simple ordered structures. In particular, the relative stability of the fluorite vs. the rutile structures is not extremely sensitive to the relative ordering of the metal atoms.

\begin{figure}[tb]
    \centering
        \includegraphics[width=9cm]{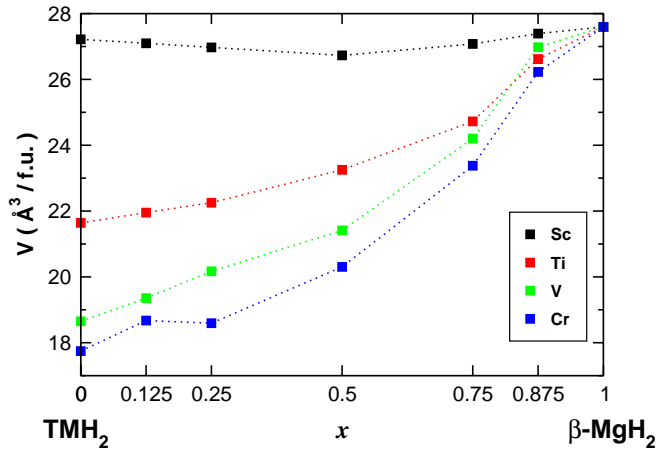}
      \caption{(Color online) The volumes per formula unit in \AA$^3$ of $\mgtmha$ in the fluorite structure, as a function of the composition $x$ for TM $=$ Sc, Ti, V, Cr (from top to bottom).}
    \label{fig:cellparameters}
\end{figure}

The calculated volumes $V({\mgtmha})$ of the fluorite structures, normalized per formula unit
(f.u.), are shown in Fig.~\ref{fig:cellparameters}. The volumes of a few bulk compounds and
compositions can be extracted from experimental data, thus providing a check on the calculations.
Interpolating the results in Ref.~\onlinecite{latroche2006csm} gives
$V(\mathrm{Mg}_{0.65}\mathrm{Sc}_{0.35}\mathrm{H}_{2})=27.5$ \AA$^3$, whereas the interpolated
calculated value from Fig.~\ref{fig:cellparameters} is 26.9 \AA$^3$.
Ref.~\onlinecite{ronnebro2005hsa} gives
$V(\mathrm{Mg}_{0.875}\mathrm{Ti}_{0.125}\mathrm{H}_{2})=27.3$ \AA$^3$, compared to the calculated
value 26.6 \AA$^3$. These differences between experimental and calculated volumes are consistent
with the differences between experimental and calculated lattice parameters of the simple
hydrides, see Table~\ref{BHstr}. Experiments on thin films give
$V(\mathrm{Mg}_{0.7}\mathrm{Ti}_{0.3}\mathrm{H}_{2})=26.4$ \AA$^3$, if one assumes cubic symmetry. \cite{borsa2007soa}   The interpolated computational value is 24.4 \AA$^3$. The difference between
these values is not excessively large, but considering the smaller difference found for the bulk
composition $\mathrm{Mg}_{0.875}\mathrm{Ti}_{0.125}\mathrm{H}_{2}$, it may suggest that the
structure of thin films is slightly different from that of bulk.

The trends observed in Fig.~\ref{fig:cellparameters} can be interpreted straightforwardly. The
cell volumes of $\alpha$-$\sch$ and the cubic $\beta$-$\mgh$ structure \cite{vajeeston2002pis} are within 1.4\% of one another,
which explains why the volumes calculated for $\mgscha$ only weakly depend on the composition $x$.
The cell volumes of the other $\tmh$ are smaller, hence one expects the volumes $V({\mgtmha})$ to increase
with $x$. At fixed composition $x$, the volumes $V({\mgtmha})$ decrease along the series Sc, Ti,
V, Cr, as the atomic volumes of the TMs decrease correspondingly. According to Zen's law of additive
volumes one would expect \cite{hafner1985nvs}
\begin{equation}
V({\mgtmha}) = xV({\beta\textrm{-}\mgh}) + (1-x)V({\tmh}).
\end{equation}
The curves shown in Fig.~\ref{fig:cellparameters} deviate slightly, but distinctly, from straight
lines, with a maximum deviation of $\sim 5$~\%. This deviation is consistent with the experimental
observations on $\mgtiha$. \cite{vermeulen2006hsm,borsa2007soa} It is also observed in simple metal alloys. \cite{hafner1985nvs}

\begin{figure}[tb]
    \centering
        \includegraphics[width=9cm]{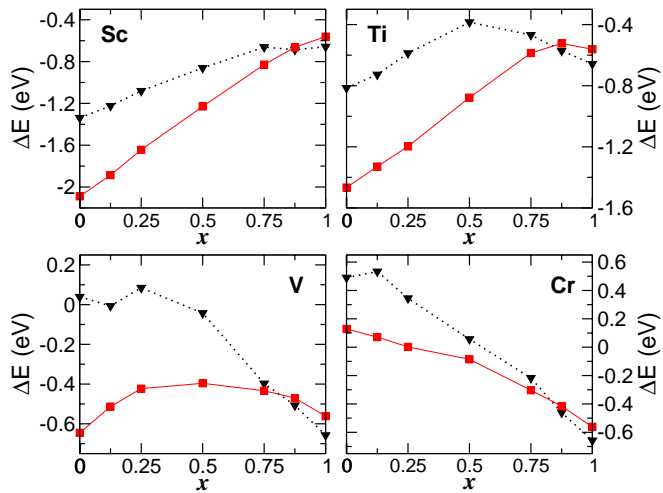}
      \caption{(Color online) The formation enthalpy (per formula unit) of the $\mgtmha$ compounds as obtained from spin polarized calculations. The values for the fluorite and rutile structures are represented by squares (solid lines) and triangles (dashed lines), respectively.}
    \label{fig:deltaH}
\end{figure}

The calculated formation enthalpies of $\mgtmha$ are shown in Fig.~\ref{fig:deltaH}.
Clearly for all TMs the fluorite structure is more stable than the rutile structure for all $x$
smaller than a critical value, $x_c$. The critical composition $x_c$ at which the rutile structure
becomes more stable can be guessed by interpolation and is in the range $x_c\approx 0.8$-0.85 for
all TMs. The fact that the critical composition is fairly high might be guessed from the energies of the fluorite and rutile structures of the simple hydrides. As a
first estimate of the critical composition $x_c$ below which the fluorite structure is stable one
may try a linear interpolation between the pure $\tmh$ compounds, $x=0$, and $\mgh$, $x=1$. The
rutile structure of $\tmh$ is more unstable than the fluorite structure by $\Delta(\tmh)=$ 0.75,
0.65, 0.68, 0.36 eV/f.u. for TM $=$ Sc, Ti, V, Cr, respectively. The difference in formation
enthalpy between the $\alpha$ (rutile) and $\beta$ (cubic) phases of $\mgh$ is $\Delta(\mgh)=0.10$
eV/f.u. Linear interpolation then gives $x_c=\Delta(\tmh)/(\Delta(\tmh)+\Delta(\mgh))$, which
results in $x_c\approx0.9$ for Sc, Ti and V, and $x_c=0.8$ for Cr. Whereas these values may seem a good first guess, they are somewhat too high as compared to the crossing points $x_c$ observed in
Fig.~\ref{fig:deltaH}. Moreover, as this figure shows, in particular the curves for Ti and V are far from linear, so the results for the linear interpolation may be somewhat fortuitous.

Experimental results indicate that Mg$_{0.7}$Ti$_{0.3}$H$_2$ has the fluorite structure and Mg$_{0.9}$Ti$_{0.1}$H$_2$ has the rutile structure. \cite{borsa2007soa} This agrees with the results shown in Fig.~\ref{fig:binarydihydrides}, where a fluorite to rutile phase transition takes place at the composition $x=x_c\approx0.83$. It has been suggested that the fluorite structure allows for a much faster kinetics of hydrogen loading and unloading. Experimentally it has been observed that the dehydrogenation kinetics of $\mgsca$ and $\mgtia$ becomes markedly slower if $x\gtrsim x_0 = 0.8$ . The results shown in Fig.~\ref{fig:binarydihydrides} suggest that $x_0 = x_c$, i.e. the composition at which the phase transition between fluorite and rutile structures takes place.

The dehydrogenation kinetics of $\mgtia$ also becomes gradually slower with
decreasing $x$, for $x<x_c$, i.e. where the compound remains in the fluorite structure, although
it is still faster than for $x>x_c$. \cite{kalisvaart2006ehs,vermeulen2006hsm} Whereas kinetic studies
are beyond the scope of the present paper, we speculate that a volume effect might play a role
here. In the fluorite structure the hydrogen atoms occupy interstitial positions close to the
tetrahedral sites. Diffusion of hydrogen atoms is likely to take place via other interstitial
sites such as the octahedral sites. The smaller the volume, the shorter the distance between such
sites and the occupied positions, or in other words, the shorter the distance between a diffusing
hydrogen atom and other hydrogen atoms in the lattice. This may increase the barrier for
diffusion. As the volume of $\mgva$ and $\mgcra$ is smaller than that of $\mgsca$ and $\mgtia$ (at
the same composition $x$), this might also explain why the dehydrogenation kinetics of the former
compounds is much slower. \cite{niessen2005ehs} We note that the smaller volume of $\mgva$ and
$\mgcra$ is accompanied by a distortion of the structures consistent with the limited space
available to accommodate the hydrogen atoms. For instance, in $\mgvhb$ and $\mgcrhb$ the hydrogen
atoms are displaced considerably from the tetrahedral positions, and the coordination number of V
and Cr (by hydrogen) is 7, instead of 8 as in case of a perfect fluorite structure.

\subsection{Electronic structure}
\begin{figure}[tb]
    \centering
        \includegraphics[width=9.5cm]{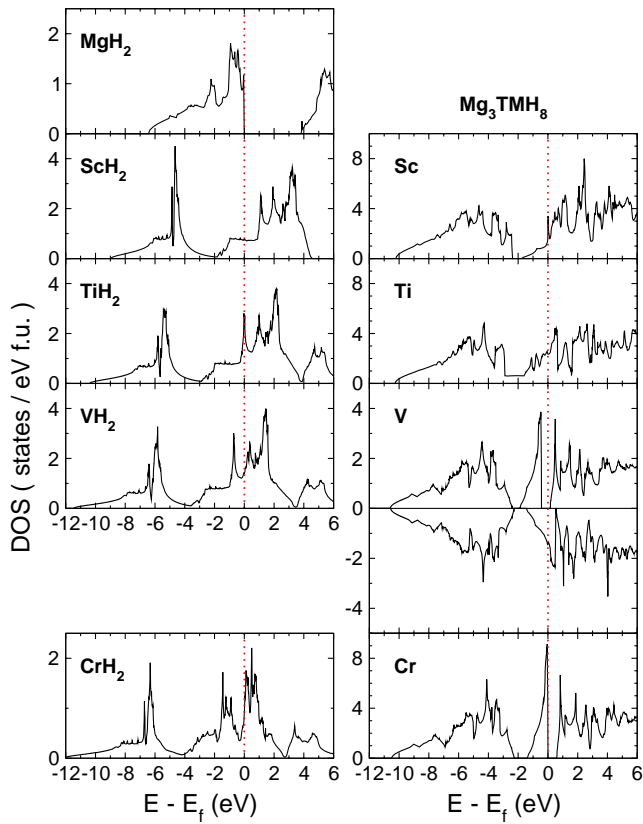}
      \caption{(Color online) Densities of states of $\mgh$ and $\tmh$ (left column) and of $\mgtmhb$ (right column) for TM $=$ Sc, Ti , V, Cr. The shaded areas give the projected densities of states on the TM $d$ orbitals. For $\crh$ the nonmagnetic DOS is given for simplicity reasons; $\crh$ is antiferromagnetic (see text).}
    \label{fig:dostmh2}
\end{figure}

To analyze the electronic structure of the compounds $\mgtmha$, we start with the density of states (DOS) of the pure hydrides $\alpha$-$\mgh$ and $\tmh$ as shown in Fig.~\ref{fig:dostmh2}. The bonding in $\mgh$ is dominantly ionic; occupied hydrogen orbitals give the main
contribution to the valence states, whereas the conduction bands have a significant contribution
from the Mg orbitals. \cite{vansetten2007esa} As usual, ionic bonding between main group elements
results in an insulator with a large band gap. In contrast, the transition metal
dihydrides are metallic, as demonstrated by Fig.~\ref{fig:dostmh2}. The peak in the DOS at low
energy, i.e. between $-9$ and $-2$ eV in $\sch$ to between $-12$ and $-4$ eV in $\crh$, is dominated by hydrogen
states. The broad peak around the Fermi level consists of transition metal $d$-states. It suggests that bonding in $\tmh$ is at least partially ionic. The TM $s$-electrons are transferred to the H atoms, whereas the $d$-electrons largely remain on the TM atoms. The DOSs of $\tmh$, TM$=$Sc, Ti, V, Cr, are very similar in shape. As the number of $d$-electrons increases from one in Sc to four in Cr, the Fermi level moves up the $d$-band in this series. As the DOS at the Fermi level increases, it enhances the probability of a magnetic instability. Indeed we find $\crh$ to be antiferromagnetic with a magnetic moment of $1.5\mu_B$ on the Cr atoms. The antiferromagnetic ordering is $51$ meV/f.u. more stable than the ferromagnetic ordering, which is $7$ meV/f.u. more stable than the non-polarized solution. In the other $\tmh$ we do not find magnetic effects.

Additional information on the type of bonding can be obtained from a Bader charge analysis. \cite{henkelman2006far}      In $\alpha$-$\mgh$ the Bader charges are $Q_\mathrm{Mg}=+1.59e$ (and
$Q_\mathrm{H}=-0.80e$, since the compound is neutral), which confirms that this compound is dominantly ionic. The results for $\tmh$
are shown in Table~\ref{table:badertmh2}. They indicate that the ionicity in $\sch$ is
comparable to that in $\mgh$. Furthermore, the ionicity decreases along the series Sc, Ti, V and $\crh$.
Comparison to Table~\ref{BHstr} shows that the decrease in ionicity correlates with a decrease in
formation enthalpy.

These results on the simple hydrides help us to analyze the electronic structure and bonding in
$\mgtmha$. We show results for the fluorite structure only, since that is the more stable structure over most of the composition range. As an example, Fig.~\ref{fig:dostmh2} shows the calculated DOSs of $\mgtmhb$. One can qualitatively interpret these DOSs as a superposition of
the DOSs of $\mgh$ and $\tmh$. The bonding states at low energy, comprising the first broad peak in the DOS, consist mainly of filled hydrogen states. The peaks close to the Fermi level are
dominated by TM $d$-states. The Fermi level moves up the $d$-band through the series Sc, Ti, V, Cr. At higher energy we find the (unoccupied) Mg $s$ states. The basic structure of the DOSs remains the same for all compositions $\mgtmha$. As $x$ increases, the TM $d$ contribution of course decreases. In addition, the TM $d$-peak becomes narrower with increasing $x$, as the distance between the TM atoms increases.

Narrowing of the $d$ peak can give rise to magnetic instabilities. The tendency to such instabilities increases along the series Sc, Ti, V and Cr. The development of nonzero magnetic moments of course strongly depends upon the structure. Nevertheless, for $\mgscha$ and $\mgtiha$ we see a tendency to form magnetic moments on the TMs only if $x\gtrsim0.8$. For $\mgvha$ this occurs if $x\gtrsim0.5$, and for $\mgcrha$ one can find magnetic instabilities over the whole composition range. Most of the structures have a finite DOS at the Fermi level, which, might indicate a metallic behavior. One cannot conclude this on the basis of a DOS alone, however, but should also critically evaluate possible localization and on-site correlation effects. There are a few exceptions. In particular cases low spin states can be more stable, such as for $\mgcrhb$ in the fluorite structure. Cubic crystal field splitting by the hydrogens surrounding the Cr atom results in a gap between $e_g$ and $t_{2g}$ states, the $e_g$ states being lowest in energy. The latter are filled by the four $d$ electrons of Cr, which makes this particular structure insulating, see Fig.~\ref{fig:dostmh2}. The DOS of $\mgvhb$ in the fluorite structure is explained by the same mechanism. However, as V only has three $d$ electrons, each V atom obtains a magnetic moment of 1~$\mu_B$. The distance between the TM atoms is fairly large in most compositions that have nonzero magnetic moments, which suggests a small magnetic coupling between the TM atoms, a low N\'{e}el or Curie temperature, and paramagnetic behavior at room temperature. Exceptions are the Cr compounds with a substantial amount of Cr, as discussed above.

A Bader charge analysis of $\mgtmha$ can be made, similar to the simple hydrides. For all
compositions $Q_\mathrm{Mg}\approx +1.6e$, i.e. close to the value found in $\alpha$-$\mgh$. As an example the Bader charges on the TM and H atoms in $\mgtmhb$ are given in
Table~\ref{table:badertmh2}. The charge on the TM atoms decreases
along the series Sc, Ti, V, and Cr as in the simple hydrides, but compared to the latter, it is somewhat smaller on V and Cr. The charges on the H atoms in $\mgtmhb$ are roughly the
proportional average of the charges on the H atoms in $\mgh$ and $\tmh$. The charge analysis of the $\mgtmha$ compounds is consistent with the bonding picture extracted from the DOSs.

\begin{table}[tb]
\centering
\caption{Bader charge analysis of $\tmh$ and $\mgtmhb$. All charges $Q$ are given in units of $e$. }
\label{table:badertmh2}
\begin{ruledtabular}
\begin{tabular}{ccccc}
          & \multicolumn{2}{c}{$\tmh$}  &  \multicolumn{2}{c}{$\mgtmhb$}  \\
TM        & $Q_\mathrm{TM}$ ($e$)  & $Q_\mathrm{H}$ ($e$)  & $Q_\mathrm{TM}$ ($e$)  & $Q_\mathrm{H}$ ($e$) \\
\hline
Sc        &   $+1.51$  &  $-0.75$  &   $+1.57$  &  $-0.80$  \\
Ti        &   $+1.17$  &  $-0.59$  &   $+1.18$  &  $-0.76$  \\
V         &   $+1.09$  &  $-0.55$  &   $+0.98$  &  $-0.73$  \\
Cr        &   $+0.89$  &  $-0.45$  &   $+0.68$  &  $-0.69$  \\
\end{tabular}
\end{ruledtabular}
\end{table}

\section{Discussion}\label{sec:discussion}
We discuss to what extend the Mg-TM alloys are suitable as hydrogen storage materials.
The formation enthalpies of $\mgtmha$ are shown in Fig.~\ref{fig:deltaH}.  Lightweight materials require a high content of magnesium, but to have a stable fluorite structure it should not exceed the critical composition $x_c$, as discussed in Sec.~\ref{sec:forment}. We focus upon the composition $\mgtmhb$ in the following discussion. The calculated formation enthalpies are $-0.83,-0.59,-0.43$ and $-0.30$ eV/f.u. for TM $=$ Sc, Ti, V, and Cr, respectively. For applications the binding enthalpy of hydrogen in the lattice should be $\lesssim 0.4$ eV/H$_2$,\cite{zuttel2003mhs,zuttel2004hsm,schlapbach2001hsm} which indicates that the Sc and Ti compounds are too stable. The formation enthalpies of the V and Cr compounds could be in the right range. However, the parameter that is most relevant for hydrogen storage is the hydrogenation enthalpy. Assuming that the alloy does not dissociate upon dehydrogenation, the hydrogenation enthalpy corresponds the reaction
\begin{equation}
\label{hydrogenation}
\mgtma  +  \hyd(g) \longrightarrow  \mgtmha.
\end{equation}

\begin{figure}[tb]
    \centering
        \includegraphics[width=8.5cm]{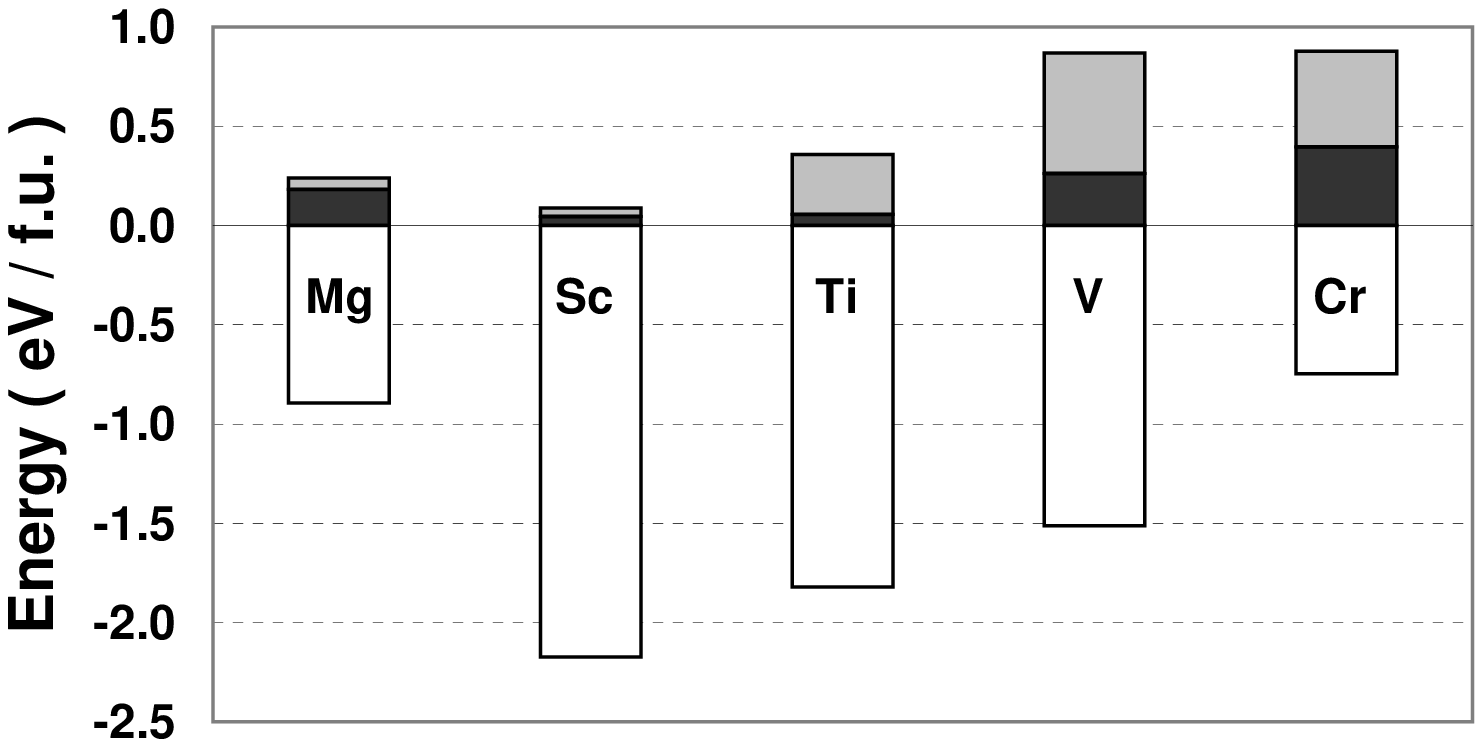}
        \includegraphics[width=8.5cm]{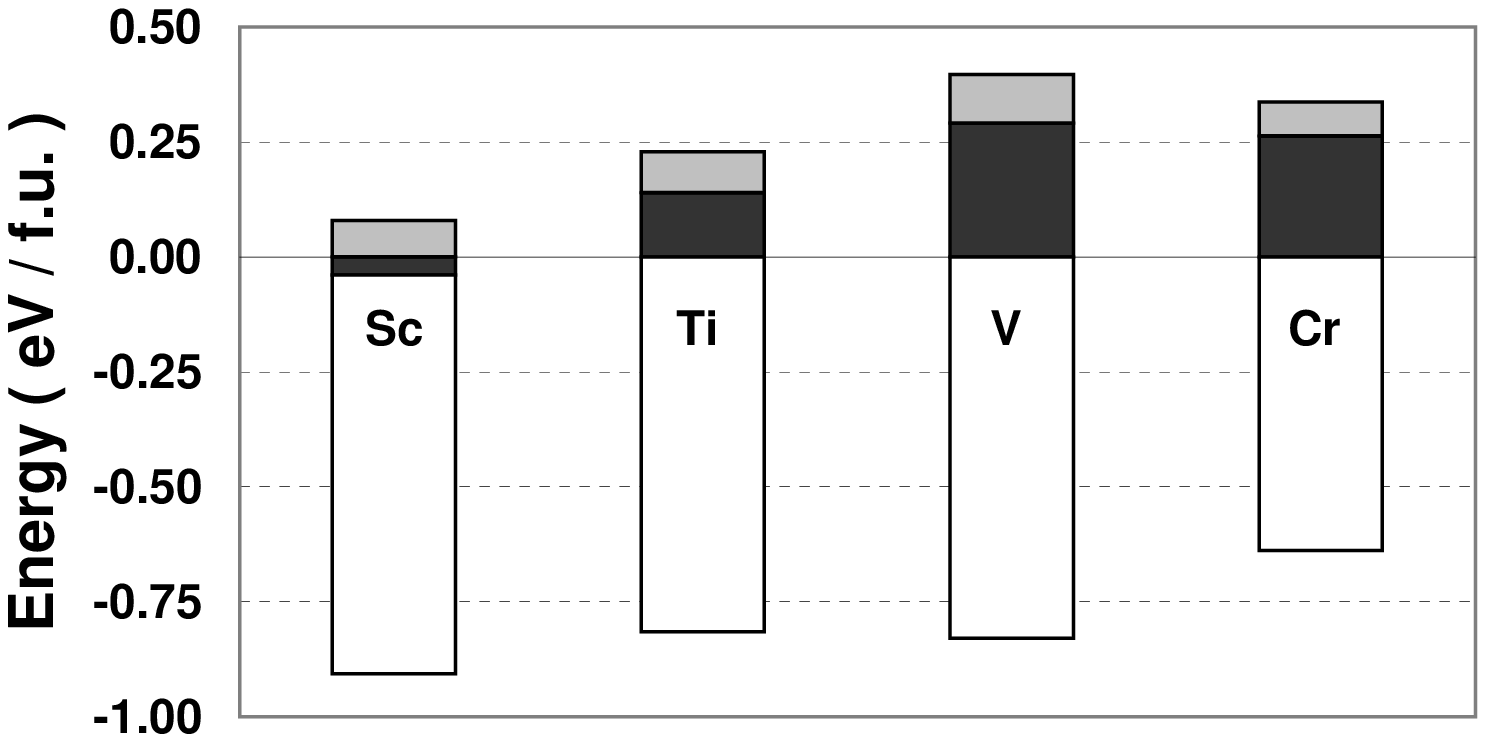}
      \caption{Decomposition of the formation energy into (i) the formation energy of the spin-polarized fcc metal $\mgtmb$ (black), (ii) the lattice expansion energy (gray), and (iii) the hydrogen insertion energy (white); (a) the simple hydrides $\mgh$, $\tmh$; (b) $\mgtmhb$.}
    \label{fig:edecomposed}
\end{figure}

To assess the hydrogenation enthalpy, one can break down the formation enthalpy associated with Eq.~(\ref{MHreaction}) into components, similar to the decomposition used in Ref.~\onlinecite{miwa2002fps}. We write the formation enthalpy as a sum of three terms. (i) The enthalpy required to make the Mg-TM alloy in the fcc structure from the elements in their most stable form. (ii) The energy required to expand the fcc lattice in order to incorporate the hydrogen atoms. (iii) The energy associated with inserting the hydrogen atoms. The results of this decomposition for $\mgtmhb$ are given in Fig.~\ref{fig:edecomposed}. To facilitate the discussion, a similar decomposition is shown for the simple hydrides, where (i) only consists of transforming the pure metal into the fcc structure. In contrast to Ref.~\onlinecite{miwa2002fps}, we use the spin-polarized fcc alloy for calculating the contributions (i) and (ii), as this will make the extraction of the hydrogenation enthalpy easier. In the cases where the magnetic moment is nonzero, we study both ferromagnetic and antiferromagnetic ordering. As for the simple hydrides, Cr compounds generally have an antiferromagnetic ordering.

The lattice expansion energy (ii) of the compounds $\mgtmhb$ is $\leq 0.1$ eV for all TMs, see Fig.~\ref{fig:edecomposed}(b). It is in fact comparable to that of pure Mg, see Fig.~\ref{fig:edecomposed}(a). At the composition $\mgtmhb$, the effect on the energy of changing the unit cell volume is dominated by Mg. For these compounds the lattice expansion only plays a minor role in the formation energy, in contrast to the simple hydrides, where the lattice expansion gives a significant contribution. The hydrogen insertion energies (iii) are also remarkably similar for the Sc, Ti, and V compounds. Again this is in sharp contrast to the corresponding energies for the simple hydrides, which strongly depend on the TM. The hydrogen insertion energies for the compounds are in fact similar to that of pure Mg. At the composition $\mgtmhb$ also this energy is then dominated by Mg. Only the compound $\mgcrhb$ has a somewhat smaller hydrogen insertion energy. The reason for this is that the energy gained by magnetic ordering of the alloy $\mgcrb$ is relatively high, as compared to the other compounds. This contribution stabilizes the alloy with respect to the hydride, which is nonmagnetic.

The formation enthalpy of the fcc alloys $\mgtmb$ (i) shows the largest variation as a function of the TM, relative to the contributions (ii) and (iii). Whereas the alloy formation energy is negative for TM $=$ Sc, indicating that this alloy is stable, it is positive for Ti, V, and Cr, meaning that these alloys are unstable. This result agrees with the experimental finding that of the Mg-TM alloys considered here, only a stable Mg-Sc alloy exists in bulk form. The substantial increase of the alloy formation enthalpy in the series Sc, Ti, V, is largely responsible for the variation of the formation energy of the corresponding hydrides $\mgtmhb$. The alloy formation energy of $\mgcrb$ is similar to that of $\mgvb$, due to a relatively high spin-polarization energy, as discussed in the previous paragraph.

The hydrogenation enthalpy according to Eq.~(\ref{hydrogenation}) can be determined by summing the contributions (ii) and (iii) of Fig.~\ref{fig:edecomposed}. Since the most stable structure of the alloys is not always the fcc structure, one should however subtract the energy required to convert the alloys from their most stable structure to an fcc structure. We find, for instance, that for $\mgtib$ the fcc structure is 0.04 eV/f.u. less stable than the hcp structure. Indeed thin film experiments on $\mgtia$ yield yield an hcp structure.\cite{kalisvaart2006ehs,borsa2007soa} For $\mgscb$ the fcc structure is more stable than the hcp structure by 0.05 eV/f.u..

The calculated hydrogenation enthalpy of $\mgscb$ is $-0.79$ eV/f.u., in good agreement with the experimental value of $-0.81$ eV/f.u..\cite{kalisvaart2006ehs} The calculated hydrogenation enthalpy of $\mgtib$ is $-0.76$ eV/f.u., which is in good agreement with the experimental value of $-0.81$ eV/f.u. of Ref. \onlinecite{gremaud2007hoc}, obtained if the thin film correction suggested there is included. These hydrogenation enthalpies are remarkably similar to that of pure Mg, strongly suggesting that alloying Mg with these TMs does not improve this energy as compared to pure Mg. The most stable structures of $\mgtmb$, TM $=$ V, Cr, are not known, but judging from the Sc and Ti compounds the energy difference between the fcc and the most stable structures will be small.
Neglecting this energy difference upper bounds for the hydrogenation enthalpies of $\mgvb$ and $\mgcrb$ are $-0.72$ and $-0.57$ eV/f.u., respectively. Again this indicates that alloying Mg with these TMs does not improve the hydrogenation enthalpy substantially.

\section{Summary}\label{sec:summary}
In summary, we have studied the structure and stability of $\mgtmha$, TM = Sc, Ti, V, Cr, compounds by first-principles calculations. We find that for $x < x_c \approx 0.8$ the fluorite structure is more stable than the rutile structure, whereas for $x > x_c$ the rutile structure is more stable. This phase transition correlates with the observed slowing down of the (de)hydrogenation kinetics in these compounds if $x$ exceeds the critical composition $x_c$. The density of states of these compounds is characterized by the valence bands being dominated by contributions from the hydrogen atoms, wherease the TMs have partially occupied $d$ states around the Fermi level. As $x$ increases and/or one moves down the TM series, the tendency to magnetic instabilities increases. The formation enthalpy of Mg$_x$TM$_{1-x}$H$_2$ can be tuned over a substantial range, i.e. 0-2 eV/f.u., by varying TM and $x$. To a large part this reflects the variation of the formation enthalpy of the alloy $\mgtma$, however. Assuming that the alloys do not decompose upon dehydrogenation, the hydrogenation enthalpy then shows much less variation. For compounds with a high magnesium content ($x=0.75$) it is close to that of pure Mg.

\section*{ACKNOWLEDGMENTS}
This work is part of the research programs of ``Advanced Chemical Technologies for Sustainability (ACTS)'' and the ``Stichting voor Fundamenteel Onderzoek der Materie (FOM)''. The use of supercomputer facilities was sponsored by the ``Stichting Nationale Computerfaciliteiten (NCF)''. These institutions are financially supported by ``Nederlandse Organisatie voor Wetenschappelijk Onderzoek (NWO)''.


\begin{thebibliography}{57}
\expandafter\ifx\csname natexlab\endcsname\relax\def\natexlab#1{#1}\fi
\expandafter\ifx\csname bibnamefont\endcsname\relax
  \def\bibnamefont#1{#1}\fi
\expandafter\ifx\csname bibfnamefont\endcsname\relax
  \def\bibfnamefont#1{#1}\fi
\expandafter\ifx\csname citenamefont\endcsname\relax
  \def\citenamefont#1{#1}\fi
\expandafter\ifx\csname url\endcsname\relax
  \def\url#1{\texttt{#1}}\fi
\expandafter\ifx\csname urlprefix\endcsname\relax\def\urlprefix{URL }\fi
\providecommand{\bibinfo}[2]{#2}
\providecommand{\eprint}[2][]{\url{#2}}

\bibitem[{coo()}]{coontz2004nss}
\bibinfo{note}{See the special issue \textit{Toward a Hydrogen Economy}, by R.
  Coontz and B. Hanson, Science \textbf{305}, 957 (2004)}.

\bibitem[{\citenamefont{Z{\"u}ttel}(2003)}]{zuttel2003mhs}
\bibinfo{author}{\bibfnamefont{A.}~\bibnamefont{Z{\"u}ttel}},
  \bibinfo{journal}{Materials Today} \textbf{\bibinfo{volume}{6}},
  \bibinfo{pages}{24} (\bibinfo{year}{2003}).

\bibitem[{\citenamefont{Z{\"u}ttel}(2004)}]{zuttel2004hsm}
\bibinfo{author}{\bibfnamefont{A.}~\bibnamefont{Z{\"u}ttel}},
  \bibinfo{journal}{Naturwissenschaften} \textbf{\bibinfo{volume}{91}},
  \bibinfo{pages}{157} (\bibinfo{year}{2004}).

\bibitem[{\citenamefont{Schlapbach and Z{\"u}ttel}(2001)}]{schlapbach2001hsm}
\bibinfo{author}{\bibfnamefont{L.}~\bibnamefont{Schlapbach}} \bibnamefont{and}
  \bibinfo{author}{\bibfnamefont{A.}~\bibnamefont{Z{\"u}ttel}},
  \bibinfo{journal}{Nature} \textbf{\bibinfo{volume}{414}},
  \bibinfo{pages}{353} (\bibinfo{year}{2001}).

\bibitem[{\citenamefont{Bogdanovic et~al.}(2003)\citenamefont{Bogdanovic,
  Felderhoff, Kaskel, Pommerin, Schlichte, and Schuth}}]{bogdanovic2003ihs}
\bibinfo{author}{\bibfnamefont{B.}~\bibnamefont{Bogdanovic}},
  \bibinfo{author}{\bibfnamefont{M.}~\bibnamefont{Felderhoff}},
  \bibinfo{author}{\bibfnamefont{S.}~\bibnamefont{Kaskel}},
  \bibinfo{author}{\bibfnamefont{A.}~\bibnamefont{Pommerin}},
  \bibinfo{author}{\bibfnamefont{K.}~\bibnamefont{Schlichte}},
  \bibnamefont{and} \bibinfo{author}{\bibfnamefont{F.}~\bibnamefont{Schuth}},
  \bibinfo{journal}{Adv. Mater.} \textbf{\bibinfo{volume}{15}},
  \bibinfo{pages}{1012} (\bibinfo{year}{2003}).

\bibitem[{\citenamefont{Stampfer~Jr et~al.}(1960)\citenamefont{Stampfer~Jr,
  Holley~Jr, and Suttle}}]{stampferjr1960mhs}
\bibinfo{author}{\bibfnamefont{J.}~\bibnamefont{Stampfer~Jr}},
  \bibinfo{author}{\bibfnamefont{C.}~\bibnamefont{Holley~Jr}},
  \bibnamefont{and} \bibinfo{author}{\bibfnamefont{J.}~\bibnamefont{Suttle}},
  \bibinfo{journal}{J. Am. Chem. Soc.} \textbf{\bibinfo{volume}{82}},
  \bibinfo{pages}{3504} (\bibinfo{year}{1960}).

\bibitem[{\citenamefont{Huot et~al.}(2001)\citenamefont{Huot, Liang, and
  Schulz}}]{huot2001mam}
\bibinfo{author}{\bibfnamefont{J.}~\bibnamefont{Huot}},
  \bibinfo{author}{\bibfnamefont{G.}~\bibnamefont{Liang}}, \bibnamefont{and}
  \bibinfo{author}{\bibfnamefont{R.}~\bibnamefont{Schulz}},
  \bibinfo{journal}{Appl. Phys. A: Mater. Sci. Process.}
  \textbf{\bibinfo{volume}{72}}, \bibinfo{pages}{187} (\bibinfo{year}{2001}).

\bibitem[{\citenamefont{Grochala and Edwards}(2004)}]{grochala2004tdn}
\bibinfo{author}{\bibfnamefont{W.}~\bibnamefont{Grochala}} \bibnamefont{and}
  \bibinfo{author}{\bibfnamefont{P.}~\bibnamefont{Edwards}},
  \bibinfo{journal}{Chem. Rev.} \textbf{\bibinfo{volume}{104}},
  \bibinfo{pages}{1283} (\bibinfo{year}{2004}).

\bibitem[{\citenamefont{Zaluska et~al.}(2001)\citenamefont{Zaluska, Zaluski,
  and Str{\"o}m-Olsen}}]{zaluska2001sca}
\bibinfo{author}{\bibfnamefont{A.}~\bibnamefont{Zaluska}},
  \bibinfo{author}{\bibfnamefont{L.}~\bibnamefont{Zaluski}}, \bibnamefont{and}
  \bibinfo{author}{\bibfnamefont{J.}~\bibnamefont{Str{\"o}m-Olsen}},
  \bibinfo{journal}{Appl. Phys. A: Mater. Sci. Process.}
  \textbf{\bibinfo{volume}{72}}, \bibinfo{pages}{157} (\bibinfo{year}{2001}).

\bibitem[{\citenamefont{Dornheim et~al.}(2006)\citenamefont{Dornheim, Eigen,
  Barkhordarian, Klassen, and Bormann}}]{dornheim2006ths}
\bibinfo{author}{\bibfnamefont{M.}~\bibnamefont{Dornheim}},
  \bibinfo{author}{\bibfnamefont{N.}~\bibnamefont{Eigen}},
  \bibinfo{author}{\bibfnamefont{G.}~\bibnamefont{Barkhordarian}},
  \bibinfo{author}{\bibfnamefont{T.}~\bibnamefont{Klassen}}, \bibnamefont{and}
  \bibinfo{author}{\bibfnamefont{R.}~\bibnamefont{Bormann}},
  \bibinfo{journal}{Adv. Eng. Mater.} \textbf{\bibinfo{volume}{8}}
  (\bibinfo{year}{2006}).

\bibitem[{\citenamefont{Li et~al.}(2007)\citenamefont{Li, Li, Ma, and
  Chen}}]{li2007mne}
\bibinfo{author}{\bibfnamefont{W.}~\bibnamefont{Li}},
  \bibinfo{author}{\bibfnamefont{C.}~\bibnamefont{Li}},
  \bibinfo{author}{\bibfnamefont{H.}~\bibnamefont{Ma}}, \bibnamefont{and}
  \bibinfo{author}{\bibfnamefont{J.}~\bibnamefont{Chen}}, \bibinfo{journal}{J.
  Am. Chem. Soc.} \textbf{\bibinfo{volume}{129}}, \bibinfo{pages}{6710}
  (\bibinfo{year}{2007}).

\bibitem[{\citenamefont{Wagemans et~al.}(2005)\citenamefont{Wagemans, van
  Lenthe, de~Jongh, van Dillen, and de~Jong}}]{wagemans2005hsm}
\bibinfo{author}{\bibfnamefont{R.}~\bibnamefont{Wagemans}},
  \bibinfo{author}{\bibfnamefont{J.}~\bibnamefont{van Lenthe}},
  \bibinfo{author}{\bibfnamefont{P.}~\bibnamefont{de~Jongh}},
  \bibinfo{author}{\bibfnamefont{A.}~\bibnamefont{van Dillen}},
  \bibnamefont{and} \bibinfo{author}{\bibfnamefont{K.}~\bibnamefont{de~Jong}},
  \bibinfo{journal}{J. Am. Chem. Soc.} \textbf{\bibinfo{volume}{127}},
  \bibinfo{pages}{16675} (\bibinfo{year}{2005}).

\bibitem[{\citenamefont{Pelletier et~al.}(2001)\citenamefont{Pelletier, Huot,
  Sutton, Schulz, Sandy, Lurio, and Mochrie}}]{pelletier2001hdm}
\bibinfo{author}{\bibfnamefont{J.}~\bibnamefont{Pelletier}},
  \bibinfo{author}{\bibfnamefont{J.}~\bibnamefont{Huot}},
  \bibinfo{author}{\bibfnamefont{M.}~\bibnamefont{Sutton}},
  \bibinfo{author}{\bibfnamefont{R.}~\bibnamefont{Schulz}},
  \bibinfo{author}{\bibfnamefont{A.}~\bibnamefont{Sandy}},
  \bibinfo{author}{\bibfnamefont{L.}~\bibnamefont{Lurio}}, \bibnamefont{and}
  \bibinfo{author}{\bibfnamefont{S.}~\bibnamefont{Mochrie}},
  \bibinfo{journal}{Phys. Rev. B} \textbf{\bibinfo{volume}{63}},
  \bibinfo{pages}{52103} (\bibinfo{year}{2001}).

\bibitem[{\citenamefont{von Zeppelin et~al.}(2002)\citenamefont{von Zeppelin,
  Reule, and Hirscher}}]{vonzeppelin2002hdk}
\bibinfo{author}{\bibfnamefont{F.}~\bibnamefont{von Zeppelin}},
  \bibinfo{author}{\bibfnamefont{H.}~\bibnamefont{Reule}}, \bibnamefont{and}
  \bibinfo{author}{\bibfnamefont{M.}~\bibnamefont{Hirscher}},
  \bibinfo{journal}{J. Alloys Compd.} \textbf{\bibinfo{volume}{330}},
  \bibinfo{pages}{723} (\bibinfo{year}{2002}).

\bibitem[{\citenamefont{Yao et~al.}(2006)\citenamefont{Yao, Wu, Du, Lu, Cheng,
  Smith, Zou, and He}}]{yao2006:jpcb}
\bibinfo{author}{\bibfnamefont{X.}~\bibnamefont{Yao}},
  \bibinfo{author}{\bibfnamefont{C.}~\bibnamefont{Wu}},
  \bibinfo{author}{\bibfnamefont{A.}~\bibnamefont{Du}},
  \bibinfo{author}{\bibfnamefont{G.~Q.} \bibnamefont{Lu}},
  \bibinfo{author}{\bibfnamefont{H.}~\bibnamefont{Cheng}},
  \bibinfo{author}{\bibfnamefont{S.~C.} \bibnamefont{Smith}},
  \bibinfo{author}{\bibfnamefont{J.}~\bibnamefont{Zou}}, \bibnamefont{and}
  \bibinfo{author}{\bibfnamefont{Y.}~\bibnamefont{He}}, \bibinfo{journal}{J.
  Phys. Chem. B} \textbf{\bibinfo{volume}{110}}, \bibinfo{pages}{11697}
  (\bibinfo{year}{2006}).

\bibitem[{\citenamefont{Notten et~al.}(2004)\citenamefont{Notten, Ouwerkerk,
  van Hal, Beelen, Keur, Zhou, and Feil}}]{notten2004hed}
\bibinfo{author}{\bibfnamefont{P.}~\bibnamefont{Notten}},
  \bibinfo{author}{\bibfnamefont{M.}~\bibnamefont{Ouwerkerk}},
  \bibinfo{author}{\bibfnamefont{H.}~\bibnamefont{van Hal}},
  \bibinfo{author}{\bibfnamefont{D.}~\bibnamefont{Beelen}},
  \bibinfo{author}{\bibfnamefont{W.}~\bibnamefont{Keur}},
  \bibinfo{author}{\bibfnamefont{J.}~\bibnamefont{Zhou}}, \bibnamefont{and}
  \bibinfo{author}{\bibfnamefont{H.}~\bibnamefont{Feil}}, \bibinfo{journal}{J.
  Power Sources} \textbf{\bibinfo{volume}{129}}, \bibinfo{pages}{45}
  (\bibinfo{year}{2004}).

\bibitem[{\citenamefont{Niessen and
  Notten}(2005{\natexlab{a}})}]{niessen2005ehs}
\bibinfo{author}{\bibfnamefont{R.}~\bibnamefont{Niessen}} \bibnamefont{and}
  \bibinfo{author}{\bibfnamefont{P.}~\bibnamefont{Notten}},
  \bibinfo{journal}{Electrochem. Solid-State Lett.}
  \textbf{\bibinfo{volume}{8}}, \bibinfo{pages}{A534}
  (\bibinfo{year}{2005}{\natexlab{a}}).

\bibitem[{\citenamefont{Niessen and
  Notten}(2005{\natexlab{b}})}]{niessen2005hst}
\bibinfo{author}{\bibfnamefont{R.}~\bibnamefont{Niessen}} \bibnamefont{and}
  \bibinfo{author}{\bibfnamefont{P.}~\bibnamefont{Notten}},
  \bibinfo{journal}{J. Alloys Compd.} \textbf{\bibinfo{volume}{404}},
  \bibinfo{pages}{457} (\bibinfo{year}{2005}{\natexlab{b}}).

\bibitem[{\citenamefont{Kalisvaart et~al.}(2006)\citenamefont{Kalisvaart,
  Niessen, and Notten}}]{kalisvaart2006ehs}
\bibinfo{author}{\bibfnamefont{W.}~\bibnamefont{Kalisvaart}},
  \bibinfo{author}{\bibfnamefont{R.}~\bibnamefont{Niessen}}, \bibnamefont{and}
  \bibinfo{author}{\bibfnamefont{P.}~\bibnamefont{Notten}},
  \bibinfo{journal}{J. Alloys Compd.} \textbf{\bibinfo{volume}{417}},
  \bibinfo{pages}{280} (\bibinfo{year}{2006}).

\bibitem[{\citenamefont{Niessen et~al.}(2006)\citenamefont{Niessen, Vermeulen,
  and Notten}}]{niessen2006epc}
\bibinfo{author}{\bibfnamefont{R.}~\bibnamefont{Niessen}},
  \bibinfo{author}{\bibfnamefont{P.}~\bibnamefont{Vermeulen}},
  \bibnamefont{and} \bibinfo{author}{\bibfnamefont{P.}~\bibnamefont{Notten}},
  \bibinfo{journal}{Electrochim. Acta} \textbf{\bibinfo{volume}{51}},
  \bibinfo{pages}{2427} (\bibinfo{year}{2006}).

\bibitem[{\citenamefont{Vermeulen
  et~al.}(2006{\natexlab{a}})\citenamefont{Vermeulen, Niessen, and
  Notten}}]{vermeulen2006hsm}
\bibinfo{author}{\bibfnamefont{P.}~\bibnamefont{Vermeulen}},
  \bibinfo{author}{\bibfnamefont{R.~A.~H.} \bibnamefont{Niessen}},
  \bibnamefont{and} \bibinfo{author}{\bibfnamefont{P.~H.~L.}
  \bibnamefont{Notten}}, \bibinfo{journal}{Electrochem. Commun.}
  \textbf{\bibinfo{volume}{8}}, \bibinfo{pages}{27}
  (\bibinfo{year}{2006}{\natexlab{a}}).

\bibitem[{\citenamefont{Borsa et~al.}(2006)\citenamefont{Borsa, Baldi,
  Pasturel, Schreuders, Dam, Griessen, Vermeulen, and Notten}}]{borsa2006mth}
\bibinfo{author}{\bibfnamefont{D.}~\bibnamefont{Borsa}},
  \bibinfo{author}{\bibfnamefont{A.}~\bibnamefont{Baldi}},
  \bibinfo{author}{\bibfnamefont{M.}~\bibnamefont{Pasturel}},
  \bibinfo{author}{\bibfnamefont{H.}~\bibnamefont{Schreuders}},
  \bibinfo{author}{\bibfnamefont{B.}~\bibnamefont{Dam}},
  \bibinfo{author}{\bibfnamefont{R.}~\bibnamefont{Griessen}},
  \bibinfo{author}{\bibfnamefont{P.}~\bibnamefont{Vermeulen}},
  \bibnamefont{and} \bibinfo{author}{\bibfnamefont{P.}~\bibnamefont{Notten}},
  \bibinfo{journal}{Appl. Phys. Lett.} \textbf{\bibinfo{volume}{88}},
  \bibinfo{pages}{241910} (\bibinfo{year}{2006}).

\bibitem[{\citenamefont{Borsa et~al.}(2007)\citenamefont{Borsa, Gremaud, Baldi,
  Schreuders, Rector, Kooi, Vermeulen, Notten, Dam, and
  Griessen}}]{borsa2007soa}
\bibinfo{author}{\bibfnamefont{D.}~\bibnamefont{Borsa}},
  \bibinfo{author}{\bibfnamefont{R.}~\bibnamefont{Gremaud}},
  \bibinfo{author}{\bibfnamefont{A.}~\bibnamefont{Baldi}},
  \bibinfo{author}{\bibfnamefont{H.}~\bibnamefont{Schreuders}},
  \bibinfo{author}{\bibfnamefont{J.}~\bibnamefont{Rector}},
  \bibinfo{author}{\bibfnamefont{B.}~\bibnamefont{Kooi}},
  \bibinfo{author}{\bibfnamefont{P.}~\bibnamefont{Vermeulen}},
  \bibinfo{author}{\bibfnamefont{P.}~\bibnamefont{Notten}},
  \bibinfo{author}{\bibfnamefont{B.}~\bibnamefont{Dam}}, \bibnamefont{and}
  \bibinfo{author}{\bibfnamefont{R.}~\bibnamefont{Griessen}},
  \bibinfo{journal}{Phys. Rev. B} \textbf{\bibinfo{volume}{75}},
  \bibinfo{pages}{205408} (\bibinfo{year}{2007}).

\bibitem[{\citenamefont{Vermeulen et~al.}(2007)\citenamefont{Vermeulen, van
  Thiel, and Notten}}]{vermeulen2007ter}
\bibinfo{author}{\bibfnamefont{P.}~\bibnamefont{Vermeulen}},
  \bibinfo{author}{\bibfnamefont{E.}~\bibnamefont{van Thiel}},
  \bibnamefont{and} \bibinfo{author}{\bibfnamefont{P.}~\bibnamefont{Notten}},
  \bibinfo{journal}{Chem.--Eur. J.} \textbf{\bibinfo{volume}{13}},
  \bibinfo{pages}{9892} (\bibinfo{year}{2007}).

\bibitem[{\citenamefont{Kalisvaart et~al.}(2007)\citenamefont{Kalisvaart,
  Wondergem, Bakker, and Notten}}]{kalisvaart2007mtb}
\bibinfo{author}{\bibfnamefont{W.~P.} \bibnamefont{Kalisvaart}},
  \bibinfo{author}{\bibfnamefont{H.~J.} \bibnamefont{Wondergem}},
  \bibinfo{author}{\bibfnamefont{F.}~\bibnamefont{Bakker}}, \bibnamefont{and}
  \bibinfo{author}{\bibfnamefont{P.~H.~L.} \bibnamefont{Notten}},
  \bibinfo{journal}{J. Mater. Res} \textbf{\bibinfo{volume}{22}},
  \bibinfo{pages}{1640} (\bibinfo{year}{2007}).

\bibitem[{\citenamefont{Gremaud et~al.}(2007)\citenamefont{Gremaud, Broedersz,
  Borsa, Borgschulte, Mauron, Schreuders, Rector, Dam, and
  Griessen}}]{gremaud2007hoc}
\bibinfo{author}{\bibfnamefont{R.}~\bibnamefont{Gremaud}},
  \bibinfo{author}{\bibfnamefont{C.}~\bibnamefont{Broedersz}},
  \bibinfo{author}{\bibfnamefont{D.}~\bibnamefont{Borsa}},
  \bibinfo{author}{\bibfnamefont{A.}~\bibnamefont{Borgschulte}},
  \bibinfo{author}{\bibfnamefont{P.}~\bibnamefont{Mauron}},
  \bibinfo{author}{\bibfnamefont{H.}~\bibnamefont{Schreuders}},
  \bibinfo{author}{\bibfnamefont{J.}~\bibnamefont{Rector}},
  \bibinfo{author}{\bibfnamefont{B.}~\bibnamefont{Dam}}, \bibnamefont{and}
  \bibinfo{author}{\bibfnamefont{R.}~\bibnamefont{Griessen}},
  \bibinfo{journal}{Adv. Mater.} \textbf{\bibinfo{volume}{19}},
  \bibinfo{pages}{2813} (\bibinfo{year}{2007}).

\bibitem[{\citenamefont{Buschow et~al.}(1982)\citenamefont{Buschow, Bouten, and
  Miedema}}]{buschow1982hfi}
\bibinfo{author}{\bibfnamefont{K.}~\bibnamefont{Buschow}},
  \bibinfo{author}{\bibfnamefont{P.}~\bibnamefont{Bouten}}, \bibnamefont{and}
  \bibinfo{author}{\bibfnamefont{A.}~\bibnamefont{Miedema}},
  \bibinfo{journal}{Reports on Progress in Physics}
  \textbf{\bibinfo{volume}{45}}, \bibinfo{pages}{937} (\bibinfo{year}{1982}).

\bibitem[{\citenamefont{Perdew et~al.}(1996)\citenamefont{Perdew, Burke, and
  Ernzerhof}}]{perdew1996gga}
\bibinfo{author}{\bibfnamefont{J.}~\bibnamefont{Perdew}},
  \bibinfo{author}{\bibfnamefont{K.}~\bibnamefont{Burke}}, \bibnamefont{and}
  \bibinfo{author}{\bibfnamefont{M.}~\bibnamefont{Ernzerhof}},
  \bibinfo{journal}{Phys. Rev. Lett.} \textbf{\bibinfo{volume}{77}},
  \bibinfo{pages}{3865} (\bibinfo{year}{1996}).

\bibitem[{\citenamefont{Bl{\"o}chl}(1994)}]{blochl1994paw}
\bibinfo{author}{\bibfnamefont{P.}~\bibnamefont{Bl{\"o}chl}},
  \bibinfo{journal}{Phys. Rev. B} \textbf{\bibinfo{volume}{50}},
  \bibinfo{pages}{17953} (\bibinfo{year}{1994}).

\bibitem[{\citenamefont{Kresse and Joubert}(1999)}]{kresse1999upp}
\bibinfo{author}{\bibfnamefont{G.}~\bibnamefont{Kresse}} \bibnamefont{and}
  \bibinfo{author}{\bibfnamefont{D.}~\bibnamefont{Joubert}},
  \bibinfo{journal}{Phys. Rev. B} \textbf{\bibinfo{volume}{59}},
  \bibinfo{pages}{1758} (\bibinfo{year}{1999}).

\bibitem[{\citenamefont{Kresse and Hafner}(1993)}]{kresse1993aim}
\bibinfo{author}{\bibfnamefont{G.}~\bibnamefont{Kresse}} \bibnamefont{and}
  \bibinfo{author}{\bibfnamefont{J.}~\bibnamefont{Hafner}},
  \bibinfo{journal}{Phys. Rev. B} \textbf{\bibinfo{volume}{47}},
  \bibinfo{pages}{558} (\bibinfo{year}{1993}).

\bibitem[{\citenamefont{Kresse and Furthm{\"u}ller}(1996)}]{kresse1996eis}
\bibinfo{author}{\bibfnamefont{G.}~\bibnamefont{Kresse}} \bibnamefont{and}
  \bibinfo{author}{\bibfnamefont{J.}~\bibnamefont{Furthm{\"u}ller}},
  \bibinfo{journal}{Phys. Rev. B} \textbf{\bibinfo{volume}{54}},
  \bibinfo{pages}{11169} (\bibinfo{year}{1996}).

\bibitem[{\citenamefont{Methfessel and Paxton}(1989)}]{methfessel1989hps}
\bibinfo{author}{\bibfnamefont{M.}~\bibnamefont{Methfessel}} \bibnamefont{and}
  \bibinfo{author}{\bibfnamefont{A.}~\bibnamefont{Paxton}},
  \bibinfo{journal}{Phys. Rev. B} \textbf{\bibinfo{volume}{40}},
  \bibinfo{pages}{3616} (\bibinfo{year}{1989}).

\bibitem[{\citenamefont{Bl{\"o}chl et~al.}(1994)\citenamefont{Bl{\"o}chl,
  Jepsen, and Andersen}}]{blochl1994itm}
\bibinfo{author}{\bibfnamefont{P.}~\bibnamefont{Bl{\"o}chl}},
  \bibinfo{author}{\bibfnamefont{O.}~\bibnamefont{Jepsen}}, \bibnamefont{and}
  \bibinfo{author}{\bibfnamefont{O.}~\bibnamefont{Andersen}},
  \bibinfo{journal}{Phys. Rev. B} \textbf{\bibinfo{volume}{49}},
  \bibinfo{pages}{16223} (\bibinfo{year}{1994}).

\bibitem[{\citenamefont{Huber and Herzberg}(1979)}]{hubher1979cdm}
\bibinfo{author}{\bibfnamefont{K.}~\bibnamefont{Huber}} \bibnamefont{and}
  \bibinfo{author}{\bibfnamefont{G.}~\bibnamefont{Herzberg}},
  \emph{\bibinfo{title}{{Molecular Spectra and Molecular Structure. IV.
  Constants of Diatomic Molecules}}} (\bibinfo{publisher}{Van Nostrand Reinhold
  Co.}, \bibinfo{year}{1979}).

\bibitem[{\citenamefont{Cox et~al.}(1989)\citenamefont{Cox, Wagman, and
  V.A.CODATA}}]{codata1989kvt}
\bibinfo{author}{\bibfnamefont{J.}~\bibnamefont{Cox}},
  \bibinfo{author}{\bibfnamefont{D.}~\bibnamefont{Wagman}}, \bibnamefont{and}
  \bibinfo{author}{\bibfnamefont{M.}~\bibnamefont{V.A.CODATA}},
  \emph{\bibinfo{title}{{Key Values for Thermodynamics}}}
  (\bibinfo{publisher}{Hemisphere}, \bibinfo{year}{1989}).

\bibitem[{\citenamefont{Bortz et~al.}(1999)\citenamefont{Bortz, Bertheville,
  Bottger, and Yvon}}]{bortz1999shp}
\bibinfo{author}{\bibfnamefont{M.}~\bibnamefont{Bortz}},
  \bibinfo{author}{\bibfnamefont{B.}~\bibnamefont{Bertheville}},
  \bibinfo{author}{\bibfnamefont{G.}~\bibnamefont{Bottger}}, \bibnamefont{and}
  \bibinfo{author}{\bibfnamefont{K.}~\bibnamefont{Yvon}}, \bibinfo{journal}{J.
  Alloys Compd.} \textbf{\bibinfo{volume}{287}}, \bibinfo{pages}{L4}
  (\bibinfo{year}{1999}).

\bibitem[{\citenamefont{Mueller et~al.}(1968)\citenamefont{Mueller, Blackledge,
  and Libowitz}}]{mueller1968mh}
\bibinfo{author}{\bibfnamefont{W.}~\bibnamefont{Mueller}},
  \bibinfo{author}{\bibfnamefont{J.}~\bibnamefont{Blackledge}},
  \bibnamefont{and} \bibinfo{author}{\bibfnamefont{G.}~\bibnamefont{Libowitz}},
  \emph{\bibinfo{title}{{Metal Hydrides}}} (\bibinfo{publisher}{Academic Press,
  New York}, \bibinfo{year}{1968}).

\bibitem[{\citenamefont{Villars et~al.}(1991)\citenamefont{Villars, Calvert,
  and Pearson}}]{villars1991psh}
\bibinfo{author}{\bibfnamefont{P.}~\bibnamefont{Villars}},
  \bibinfo{author}{\bibfnamefont{L.}~\bibnamefont{Calvert}}, \bibnamefont{and}
  \bibinfo{author}{\bibfnamefont{W.}~\bibnamefont{Pearson}},
  \emph{\bibinfo{title}{{Pearson's Handbook of Crystallographic Data for
  Intermetallic Phases}}} (\bibinfo{publisher}{ASM International, Materials
  Park, OH, USA}, \bibinfo{year}{1991}).

\bibitem[{\citenamefont{Snavely and Vaughan}(1949)}]{snavely1949ucd}
\bibinfo{author}{\bibfnamefont{C.}~\bibnamefont{Snavely}} \bibnamefont{and}
  \bibinfo{author}{\bibfnamefont{D.}~\bibnamefont{Vaughan}},
  \bibinfo{journal}{J. Am. Chem. Soc.} \textbf{\bibinfo{volume}{71}},
  \bibinfo{pages}{313} (\bibinfo{year}{1949}).

\bibitem[{\citenamefont{Latroche et~al.}(2006)\citenamefont{Latroche,
  Kalisvaart, and Notten}}]{latroche2006csm}
\bibinfo{author}{\bibfnamefont{M.}~\bibnamefont{Latroche}},
  \bibinfo{author}{\bibfnamefont{P.}~\bibnamefont{Kalisvaart}},
  \bibnamefont{and} \bibinfo{author}{\bibfnamefont{P.}~\bibnamefont{Notten}},
  \bibinfo{journal}{J. Solid State Chem.} \textbf{\bibinfo{volume}{179}},
  \bibinfo{pages}{3024} (\bibinfo{year}{2006}).

\bibitem[{\citenamefont{Magusin et~al.}(2008)\citenamefont{Magusin, Kalisvaart,
  Notten, and van Santen}}]{magusin2008hsa}
\bibinfo{author}{\bibfnamefont{P.}~\bibnamefont{Magusin}},
  \bibinfo{author}{\bibfnamefont{W.}~\bibnamefont{Kalisvaart}},
  \bibinfo{author}{\bibfnamefont{P.}~\bibnamefont{Notten}}, \bibnamefont{and}
  \bibinfo{author}{\bibfnamefont{R.}~\bibnamefont{van Santen}},
  \bibinfo{journal}{Chem. Phys. Lett.}  (\bibinfo{year}{2008}).

\bibitem[{\citenamefont{Vermeulen
  et~al.}(2006{\natexlab{b}})\citenamefont{Vermeulen, Niessen, Borsa, Dam,
  Griessen, and Notten}}]{vermeulen2006edt}
\bibinfo{author}{\bibfnamefont{P.}~\bibnamefont{Vermeulen}},
  \bibinfo{author}{\bibfnamefont{R.}~\bibnamefont{Niessen}},
  \bibinfo{author}{\bibfnamefont{D.}~\bibnamefont{Borsa}},
  \bibinfo{author}{\bibfnamefont{B.}~\bibnamefont{Dam}},
  \bibinfo{author}{\bibfnamefont{R.}~\bibnamefont{Griessen}}, \bibnamefont{and}
  \bibinfo{author}{\bibfnamefont{P.}~\bibnamefont{Notten}},
  \bibinfo{journal}{Electrochem. Solid-State Lett.}
  \textbf{\bibinfo{volume}{9}}, \bibinfo{pages}{A520}
  (\bibinfo{year}{2006}{\natexlab{b}}).

\bibitem[{\citenamefont{Liang et~al.}(1999)\citenamefont{Liang, Huot, Boily,
  Van~Neste, and Schulz}}]{liang1999cet}
\bibinfo{author}{\bibfnamefont{G.}~\bibnamefont{Liang}},
  \bibinfo{author}{\bibfnamefont{J.}~\bibnamefont{Huot}},
  \bibinfo{author}{\bibfnamefont{S.}~\bibnamefont{Boily}},
  \bibinfo{author}{\bibfnamefont{A.}~\bibnamefont{Van~Neste}},
  \bibnamefont{and} \bibinfo{author}{\bibfnamefont{R.}~\bibnamefont{Schulz}},
  \bibinfo{journal}{J. Alloys Compd.} \textbf{\bibinfo{volume}{292}},
  \bibinfo{pages}{247} (\bibinfo{year}{1999}).

\bibitem[{\citenamefont{Bobet et~al.}(2000)\citenamefont{Bobet, Even, Nakamura,
  Akiba, and Darriet}}]{bobet2000sma}
\bibinfo{author}{\bibfnamefont{J.}~\bibnamefont{Bobet}},
  \bibinfo{author}{\bibfnamefont{C.}~\bibnamefont{Even}},
  \bibinfo{author}{\bibfnamefont{Y.}~\bibnamefont{Nakamura}},
  \bibinfo{author}{\bibfnamefont{E.}~\bibnamefont{Akiba}}, \bibnamefont{and}
  \bibinfo{author}{\bibfnamefont{B.}~\bibnamefont{Darriet}},
  \bibinfo{journal}{J. Alloys Compd.} \textbf{\bibinfo{volume}{298}},
  \bibinfo{pages}{279} (\bibinfo{year}{2000}).

\bibitem[{\citenamefont{Liang and Schulz}(2003)}]{liang2003smt}
\bibinfo{author}{\bibfnamefont{G.}~\bibnamefont{Liang}} \bibnamefont{and}
  \bibinfo{author}{\bibfnamefont{R.}~\bibnamefont{Schulz}},
  \bibinfo{journal}{J. Mater. Sci.} \textbf{\bibinfo{volume}{38}},
  \bibinfo{pages}{1179} (\bibinfo{year}{2003}).

\bibitem[{\citenamefont{Choi et~al.}(2008)\citenamefont{Choi, Lu, Sohn, and
  Fang}}]{choi2008hsp}
\bibinfo{author}{\bibfnamefont{Y.}~\bibnamefont{Choi}},
  \bibinfo{author}{\bibfnamefont{J.}~\bibnamefont{Lu}},
  \bibinfo{author}{\bibfnamefont{H.}~\bibnamefont{Sohn}}, \bibnamefont{and}
  \bibinfo{author}{\bibfnamefont{Z.}~\bibnamefont{Fang}}, \bibinfo{journal}{J.
  Power Sources} \textbf{\bibinfo{volume}{180}}, \bibinfo{pages}{491}
  (\bibinfo{year}{2008}).

\bibitem[{\citenamefont{Kyoi et~al.}(2004{\natexlab{a}})\citenamefont{Kyoi,
  Sato, R{\"o}nnebro, Kitamura, Ueda, Ito, Katsuyama, Hara, Noreus, and
  Sakai}}]{kyoi2004ntm}
\bibinfo{author}{\bibfnamefont{D.}~\bibnamefont{Kyoi}},
  \bibinfo{author}{\bibfnamefont{T.}~\bibnamefont{Sato}},
  \bibinfo{author}{\bibfnamefont{E.}~\bibnamefont{R{\"o}nnebro}},
  \bibinfo{author}{\bibfnamefont{N.}~\bibnamefont{Kitamura}},
  \bibinfo{author}{\bibfnamefont{A.}~\bibnamefont{Ueda}},
  \bibinfo{author}{\bibfnamefont{M.}~\bibnamefont{Ito}},
  \bibinfo{author}{\bibfnamefont{S.}~\bibnamefont{Katsuyama}},
  \bibinfo{author}{\bibfnamefont{S.}~\bibnamefont{Hara}},
  \bibinfo{author}{\bibfnamefont{D.}~\bibnamefont{Noreus}}, \bibnamefont{and}
  \bibinfo{author}{\bibfnamefont{T.}~\bibnamefont{Sakai}}, \bibinfo{journal}{J.
  Alloys Compd.} \textbf{\bibinfo{volume}{372}}, \bibinfo{pages}{213}
  (\bibinfo{year}{2004}{\natexlab{a}}).

\bibitem[{\citenamefont{Kyoi et~al.}(2003)\citenamefont{Kyoi, Ronnebro,
  Kitamura, Ueda, Ito, Katsuyama, and Sakai}}]{kyoi2003fmc}
\bibinfo{author}{\bibfnamefont{D.}~\bibnamefont{Kyoi}},
  \bibinfo{author}{\bibfnamefont{E.}~\bibnamefont{Ronnebro}},
  \bibinfo{author}{\bibfnamefont{N.}~\bibnamefont{Kitamura}},
  \bibinfo{author}{\bibfnamefont{A.}~\bibnamefont{Ueda}},
  \bibinfo{author}{\bibfnamefont{M.}~\bibnamefont{Ito}},
  \bibinfo{author}{\bibfnamefont{S.}~\bibnamefont{Katsuyama}},
  \bibnamefont{and} \bibinfo{author}{\bibfnamefont{T.}~\bibnamefont{Sakai}},
  \bibinfo{journal}{J. Alloys Compd.} \textbf{\bibinfo{volume}{361}},
  \bibinfo{pages}{252} (\bibinfo{year}{2003}).

\bibitem[{\citenamefont{Kyoi et~al.}(2004{\natexlab{b}})\citenamefont{Kyoi,
  Sato, Ronnebro, Tsuji, Kitamura, Ueda, Ito, Katsuyama, Hara, Noreus
  et~al.}}]{kyoi2004nmv}
\bibinfo{author}{\bibfnamefont{D.}~\bibnamefont{Kyoi}},
  \bibinfo{author}{\bibfnamefont{T.}~\bibnamefont{Sato}},
  \bibinfo{author}{\bibfnamefont{E.}~\bibnamefont{Ronnebro}},
  \bibinfo{author}{\bibfnamefont{Y.}~\bibnamefont{Tsuji}},
  \bibinfo{author}{\bibfnamefont{N.}~\bibnamefont{Kitamura}},
  \bibinfo{author}{\bibfnamefont{A.}~\bibnamefont{Ueda}},
  \bibinfo{author}{\bibfnamefont{M.}~\bibnamefont{Ito}},
  \bibinfo{author}{\bibfnamefont{S.}~\bibnamefont{Katsuyama}},
  \bibinfo{author}{\bibfnamefont{S.}~\bibnamefont{Hara}},
  \bibinfo{author}{\bibfnamefont{D.}~\bibnamefont{Noreus}},
  \bibnamefont{et~al.}, \bibinfo{journal}{J. Alloys Compd.}
  \textbf{\bibinfo{volume}{375}}, \bibinfo{pages}{253}
  (\bibinfo{year}{2004}{\natexlab{b}}).

\bibitem[{\citenamefont{R{\"o}nnebro et~al.}(2004)\citenamefont{R{\"o}nnebro,
  Kyoi, Blomqvist, Nor{\'e}us, and Sakai}}]{ronnebro2004scm}
\bibinfo{author}{\bibfnamefont{E.}~\bibnamefont{R{\"o}nnebro}},
  \bibinfo{author}{\bibfnamefont{D.}~\bibnamefont{Kyoi}},
  \bibinfo{author}{\bibfnamefont{H.}~\bibnamefont{Blomqvist}},
  \bibinfo{author}{\bibfnamefont{D.}~\bibnamefont{Nor{\'e}us}},
  \bibnamefont{and} \bibinfo{author}{\bibfnamefont{T.}~\bibnamefont{Sakai}},
  \bibinfo{journal}{J. Alloys Compd.} \textbf{\bibinfo{volume}{368}},
  \bibinfo{pages}{279} (\bibinfo{year}{2004}).

\bibitem[{\citenamefont{R{\"o}nnebro et~al.}(2005)\citenamefont{R{\"o}nnebro,
  Kyoi, Kitano, Kitano, and Sakai}}]{ronnebro2005hsa}
\bibinfo{author}{\bibfnamefont{E.}~\bibnamefont{R{\"o}nnebro}},
  \bibinfo{author}{\bibfnamefont{D.}~\bibnamefont{Kyoi}},
  \bibinfo{author}{\bibfnamefont{A.}~\bibnamefont{Kitano}},
  \bibinfo{author}{\bibfnamefont{Y.}~\bibnamefont{Kitano}}, \bibnamefont{and}
  \bibinfo{author}{\bibfnamefont{T.}~\bibnamefont{Sakai}}, \bibinfo{journal}{J.
  Alloys Compd.} \textbf{\bibinfo{volume}{404}}, \bibinfo{pages}{68}
  (\bibinfo{year}{2005}).

\bibitem[{\citenamefont{Vajeeston et~al.}(2002)\citenamefont{Vajeeston,
  Ravindran, Kjekshus, and Fjellv{\aa}g}}]{vajeeston2002pis}
\bibinfo{author}{\bibfnamefont{P.}~\bibnamefont{Vajeeston}},
  \bibinfo{author}{\bibfnamefont{P.}~\bibnamefont{Ravindran}},
  \bibinfo{author}{\bibfnamefont{A.}~\bibnamefont{Kjekshus}}, \bibnamefont{and}
  \bibinfo{author}{\bibfnamefont{H.}~\bibnamefont{Fjellv{\aa}g}},
  \bibinfo{journal}{Phys. Rev. Lett.} \textbf{\bibinfo{volume}{89}},
  \bibinfo{pages}{175506} (\bibinfo{year}{2002}).

\bibitem[{\citenamefont{Hafner}(1985)}]{hafner1985nvs}
\bibinfo{author}{\bibfnamefont{J.}~\bibnamefont{Hafner}}, \bibinfo{journal}{J.
  Phys. F: Met. Phys.} \textbf{\bibinfo{volume}{15}}, \bibinfo{pages}{L43}
  (\bibinfo{year}{1985}).

\bibitem[{\citenamefont{van Setten et~al.}(2007)\citenamefont{van Setten, Popa,
  de~Wijs, and Brocks}}]{vansetten2007esa}
\bibinfo{author}{\bibfnamefont{M.}~\bibnamefont{van Setten}},
  \bibinfo{author}{\bibfnamefont{V.}~\bibnamefont{Popa}},
  \bibinfo{author}{\bibfnamefont{G.}~\bibnamefont{de~Wijs}}, \bibnamefont{and}
  \bibinfo{author}{\bibfnamefont{G.}~\bibnamefont{Brocks}},
  \bibinfo{journal}{Phys. Rev. B} \textbf{\bibinfo{volume}{75}},
  \bibinfo{pages}{35204} (\bibinfo{year}{2007}).

\bibitem[{\citenamefont{Henkelman et~al.}(2006)\citenamefont{Henkelman,
  Arnaldsson, and J{\'o}nsson}}]{henkelman2006far}
\bibinfo{author}{\bibfnamefont{G.}~\bibnamefont{Henkelman}},
  \bibinfo{author}{\bibfnamefont{A.}~\bibnamefont{Arnaldsson}},
  \bibnamefont{and}
  \bibinfo{author}{\bibfnamefont{H.}~\bibnamefont{J{\'o}nsson}},
  \bibinfo{journal}{Comput. Mater. Sci.} \textbf{\bibinfo{volume}{36}},
  \bibinfo{pages}{354} (\bibinfo{year}{2006}).

\bibitem[{\citenamefont{Miwa and Fukumoto}(2002)}]{miwa2002fps}
\bibinfo{author}{\bibfnamefont{K.}~\bibnamefont{Miwa}} \bibnamefont{and}
  \bibinfo{author}{\bibfnamefont{A.}~\bibnamefont{Fukumoto}},
  \bibinfo{journal}{Phys. Rev. B} \textbf{\bibinfo{volume}{65}},
  \bibinfo{pages}{155114} (\bibinfo{year}{2002}).

\end{thebibliography}

\end{document}